\documentclass[a4paper]{jpconf}
\usepackage{graphicx}
\usepackage{amsmath}
\usepackage[bookmarks=true,colorlinks=true,citecolor=blue,linkcolor=red,urlcolor=magenta]{hyperref}

\begin{document}
\title{Quantization in relativistic classical mechanics: the Stueckelberg equation, neutrino oscillation and large-scale structure of the Universe 
}

\author{V D Rusov and D S Vlasenko}

\address{Department of Theoretical and Experimental Nuclear Physics,
Odessa National Polytechnic University, 65044 Odessa, Ukraine
}

\ead{siiis@te.net.ua}

\begin{abstract}
Based on the Chetaev theorem on stable dynamical trajectories in the presence of dissipative forces, we obtain the generalized condition for stability of relativistic classical Hamiltonian systems (with an invariant evolution parameter) in the form of the Stueckelberg equation. As is known, this equation is the basis of a competing paradigm known as parametrized relativistic quantum mechanics (pRQM).

It is shown that the energy of dissipative forces, which generate the Chetaev generalized condition of stability, coincides exactly with Bohmian relativistic quantum potential. Within the framework of Bohmian RQM supplemented by the generalized Chetaev theorem and on the basis of the principle of least action for dissipative forces, we show that the squared amplitude of a wave function in the Stueckelberg equation is equivalent to the probability density function for the number of particle trajectories, relative to which the velocity and the position of the particle are not hidden parameters.

The conditions for reasonableness of trajectory interpretation of pRQM are discussed. Based on analysis of a general formalism for vacuum-flavor mixing of neutrino within the context of the standard and pRQM models we show that the corresponding expressions for the probability of transition from one neutrino flavor to another differ appreciably, but they are experimentally testable: the estimations of absolute value for neutrino mass based on modern experimental data for solar and atmospheric neutrinos show that the pRQM results have a preference. It is noted that the selection criterion of mass solutions relies on proximity between the average size of condensed neutrino clouds, which is described by the Muraki formula (29th ICRC, 2005) and depends on the neutrino mass, and the average size of typical observed void structure (dark matter + hydrogen gas), which plays the role of characteristic dimension of large-scale structure of the Universe.
\end{abstract}

\section{Introduction}

In this paper we continue to consider the question, which is formulated in \cite{bib-01,bib-02} in the following rather strict and paradoxical form: "Are the so-called quantization conditions that are imposed on the corresponding spectrum of a dynamical system possible in principle in classical mechanics, analogously to what is taking place in quantum mechanics?" As is shown in \cite{bib-01,bib-02}, the answer to this question is positive and has been given more than 70 years ago by the Russian mathematician and mechanician N.G. Chetaev in his article "On stable trajectories in dynamics" \cite{bib-03,bib-04,bib-05,bib-06}. What is more, based on the Chetaev theorem on stable dynamical trajectories in the presence of dissipative forces \cite{bib-03,bib-04,bib-05,bib-06}, we obtained the generalized condition for stability of Hamilton systems in the form of the Schr\"{o}dinger equation \cite{bib-01,bib-02}. 

At the same time, it is necessary to remind that the essential difference of the Chetaev quantum mechanics with trajectories from the known alternative quantum theories [7-16] is the unconditional identity of the probability density function of the number of particle trajectories (obtained by the principle of least action of perturbations (see (28) in \cite{bib-01,bib-02}) and the probability density function for a particle to be in a certain place of the configuration space (obtained by the Bohm-Madelung equation of continuity (see (25) in \cite{bib-01,bib-02})). This exclusively important fact emphasizes naturally the physical identity of the probabilistic and trajectory interpretations of nonrelativistic quantum mechanics.

In this paper we show that similar results can be obtained in the framework of relativistic classical mechanics. In other words, based on generalized stability conditions for stability of Hamilton relativistic systems (with invariant evolution parameter) we obtained the relativistic Schr\"{o}dinger equation or, more exactly, the so-called Stueckelberg equation, which as is known is the basis of a competing paradigm known as parametrized relativistic quantum mechanics (pRQM) \cite{bib-17}.
 
On the other hand, it is obvious that the fact of appearance of an alternative quantum theory causes immediately a number of questions connected, in the first place, with comparison between results of alternative and standard quantum theories obtained in the same experiments. 
As is exactly mentioned by Adler and Bassi \cite{bib-18}, "...but apart from this history, there is another important motivation for considering modifications of quantum theory. This is to give a quantitative meaning to experiments testing quantum theory, by having an alternative theory, making predictions that differ from those of standard quantum theory, to which these experiments can be compared".

Thus, the main purpose of the present work is, on the one hand, derivation of the Schr\"{o}dinger relativistic equation or, more exactly, so-called the Stueckelberg equation based on the generalized condition for stability of Hamilton relativistic systems (with invariant evolution parameter) and, on the other hand, quantitative comparison between results of alternative (pRQM) and standard quantum theories by analysis of general formalism for vacuum-flavor mixing of neutrino.

\section{The emergent Stueckelberg equation in relativistic classical mechanics}

In what follows, we present a generalization of the Chetaev theorem on stable trajectories in dynamics to the case of relativistic Hamiltonian $K(s,q^{i},p_{i})$, where $s$ is the evolution parameter of a system $i=0,1,2,3$. With this purpose, we consider a material system, in which $q^{i}$ and $p_{i}$  are the generalized 4-space coordinates and momenta of the holonomic system in the field of potential forces which admit a force function of the form $U(q^{i})$.

In the general case where the action $S$ is the explicit function of the evolution parameter $s$, the total integral of the Hamilton-Jacobi differential equation corresponding to a given system takes the form: 

\begin{equation}
\label{eq1}
S=f(s,q^{i};\alpha^{i} )+C,
\end{equation}

\noindent where $\alpha^{i}$ and $C$,  are arbitrary constants, and the general solution of the mechanical problem is given, according to the well-known Jacobi theorem, by the formulas: 

\begin{equation}
\label{eq2}
\beta^{i} =\partial S/\partial \alpha_{i}, ~~~ {p_{i} =\partial S/\partial q^{i} }, ~~~ {i=0,...,3},
\end{equation}

\noindent where $\beta^{i}$ are the new constants of integration. The possible motions of the mechanical system are defined by various values of the constants $\alpha^{i}$ and $\beta^{i}$.

We call a motion of the material system, whose stability will be studied, the unperturbed motion. First, we consider the stability of such motion with respect to the variables $q^{i}$ under a perturbation of only the initial values of these variables (i.e. of values of the constants $\alpha^{i}$ and $\beta^{i}$) without disturbing forces. Omitting the details of the derivation which are given in \cite{bib-03,bib-04,bib-06,bib-19}, we present the necessary Chetaev stability condition in this case as follows:

\begin{equation}
\label{eq3}
\Lambda=\frac{1}{m}\partial_{i}\partial^{i}S=0 ,
\end{equation}

We now complicate the problem. Let a really moving material system undergo the action of both the forces with force function U, theoretically considered above, and the unknown perturbation (dissipative) forces which are assumed to be potential and admit the dissipative force function $Q$. Then the real motion of the material system occurs in the field of forces with general force function $U^{*} = U+Q$. In this case, the real motion of the system does not coincide obviously at all with the theoretical one (without perturbation).

If we conserve the statement of the problem on the stability of real unperturbed motions in the theoretical field of forces with the function $U$ at a perturbation of only initial data as above, then the necessary requirement of stability in the first approximation, e.g., in form (\ref{eq3}), will not be efficient in the general case, because the new function $S$ is unknown (as well as $Q$). However, it turns out that we can determine such conditions of stability, which are implicitly independent of the form of the unknown action functions $S$ and the potential $Q$. Thus, we are based on the requirement of stability of form (\ref{eq3}), by assuming the conditions of its existence (correctness, etc.) for real motions to be satisfied. In relation (\ref{eq3}), we now replace the function $S$ by a new function $A$ defined by the equality:

\begin{equation}
\label{eq4}
\psi =A\exp (ikS),
\end{equation}

\noindent where $k$ is a constant; $A$ is a function of the generalized coordinates $q$ and the evolution parameter $s$.

The introduction of a real wave function, like the de~Broglie "pilot-wave" \cite{bib-20}, is extremely necessary from the physical viewpoint because of the following non-trivial reason. Since the dynamics of a physical system must undoubtedly conserve the Hamilton form of the equations of motion, the main "task" of such a real wave consists in the exact compensation of the action of dissipative forces, which are generated by the perturbation energy $Q$. We will show below that, in this case, such a procedure makes it possible not only to conserve the Hamilton form of the dynamics of a physical system, but allows one to determine the character of an analytic dependence of the energy $Q$ of disturbing forces on the wave function amplitude $A$. Then relation (\ref{eq4}) yields:

\begin{equation}
\label{eq5}
\partial_{j}S=\frac{1}{ik}\left( {\frac{1}{\psi 
}\partial_{j}\psi-\frac{1}{A}\partial_{j}A} \right)
\end{equation}

\noindent and, hence, relation (\ref{eq3}) looks like

\begin{equation}
\label{eq6}
\frac{1}{m}\partial_{i}\left[ \frac{1}{\psi }\partial^{i}\psi-\frac{1}{A}\partial^{i}A \right] =0.
\end{equation}

On the other hand, we can write the Hamilton-Jacobi equations for a perturbed motion in the general case where the Hamiltonian $H$ depends explicitly on the evolution parameter $s$:

\[
\frac{\partial S}{\partial s}+\frac{1}{2m}\partial_{i}S \partial^{i}S+U+Q=0,
\]

\noindent or with the help of (\ref{eq4}-\ref{eq6})

\begin{equation}
\label{eq7}
\frac{1}{2mk^2}\left( \frac{1}{\psi} 
\partial_{i}\psi-\frac{1}{A}\partial_{i}A \right) \left( \frac{1}{\psi }\partial^{i}\psi
-\frac{1}{A}\partial^{i}A \right)=\frac{\partial 
S}{\partial s}+U+Q,
\end{equation}

\noindent where $\partial S$/$\partial s $ can be determined with the help of (\ref{eq4}). Adding relations (\ref{eq6}) and (\ref{eq7}), we obtain the necessary condition of stability (in the first approximation) in the form:

\begin{equation}
\label{eq8}
  \begin{split}
    \frac{1}{2mk^2\psi }\partial_{i}\partial^{i}\psi-\frac{1}{2k^2A}\partial_{i}\partial^{i}A - \frac{1}{mk^2A}\partial_{i}A\left(\frac{1}{\psi }\partial_{i}\psi-\frac{1}{A}\partial_{i}A \right)-\\
    -\frac{1}{ikA\psi}\left[ {A\partial_{s}\psi-\psi \partial_{s}A} \right]-U-Q=0.
  \end{split}
\end{equation}

In this place, we need to use a procedure for the compensation of the action of dissipative forces, which are generated by the perturbation energy $Q$ in order to conserve the Hamilton form of the dynamics of a physical system (\ref{eq8}). It is obvious that equality (\ref{eq8}) will not contain $Q$, if the amplitude $A$ is determined from the equation:

\begin{equation}
\label{eq9}
\frac{1}{2k^2A}\partial_{i}\partial^{i}A+
\frac{i}{kA}\partial_{i}A\partial^{i}S-\frac{1}{ikA}\partial_{s}A+Q=0,
\end{equation}

\noindent which splits into two equations

\begin{equation}
\label{eq10}
Q=-\frac{1}{2mk^2A}\partial_{i}\partial^{i}A,
\end{equation}

\begin{equation}
\label{eq11}
\partial_{s}A=-\frac{1}{m}\partial_{i}A\partial^{i}S,
\end{equation}

\noindent after the separation of the real and imaginary parts. Here Q is the dissipation energy. Thus, if the properties of disturbing forces satisfy conditions (\ref{eq10}) and (\ref{eq11}), then the necessary condition of stability (\ref{eq8}) has the form of a differential equation of the "Stueckelberg" type:

\begin{equation}
\label{eq12}
\frac{i}{k}\partial_{s}\psi=-\frac{1}{2mk^2}\partial_{i}\partial^{i}\psi + U\psi ,
\end{equation}

\noindent where $q^{\mu}(s)$ gives the position of the physical system, whose possible trajectories in the 4-dimensional configuration space $q = (q_{0},,q_{3}$) are a solution of the system of the so-called governing equations \cite{bib-21}:

\begin{equation}
\label{eq13}
v_{i}=\frac{1}{m}\partial_{j} S,
\end{equation}

\noindent where $S$ is the phase of the wave function. 

In other words, we obtained the following result: Eq.~(\ref{eq3}), which corresponds to the Chetaev stability
condition, is transformed into an equation of the "Stueckelberg" type (\ref{eq12}) with the use of transformation
(\ref{eq4}). It should be noted  that in spite of similarity of the Stueckelberg and Schr\"{o}dinger equations they are different.
Evolution parameter  differs from the time \cite{bib-17,bib-22}. It is obvious that, in the class of equations of type (\ref{eq12}), the single-valued, finite, and continuous solutions for the function $\psi$ in the stationary case are admissible only for the eigenvalues
of the total energy $E$. Hence, the given stability of real motions takes place only for these values of the
total energy $E$. 

We now present a short analysis of the obtained results. It is known that, according to one of the theorems of stability theory \cite{bib-06}, only two types of forces - dissipative and gyroscopic  do not break the stability (if it is present) of a non-disturbed motion of holonomic mechanical systems. Therefore, by introducing a dissipative perturbation $Q$ into the Hamilton-Jacobi Eq.~(\ref{eq7}) and by taking simultaneously the stability condition (\ref{eq3}) into account, we reasonably expect that, under condition of the conservation of the stability of a mechanical system, it is possible to get a real functional dependence of the dissipation energy $Q$ on characteristics of the wave function $\psi$ (\ref{eq4}). Indeed, having obtained the condition for the stability of trajectories of a dynamical system in the form of an equation of the "Stueckelberg" type (\ref{eq12}), we established not only the physical sense of the perturbation energy $Q$, but  showed also that it is a function of the amplitude of the wave function $\psi$ (\ref{eq4}) and takes form (\ref{eq10}). It is a very important result. The subsequent content of the article is, as will be clear in what follows, a direct consequence of this result.
 
It is easy to show that namely the conclusion about the dissipative nature together with the simultaneous determination of a functional dependence of the energy of disturbing forces on the form (but not on the magnitude ) of the amplitude of the wave function $\psi$  allow one to generalize an equation of the "Stueckelberg" type (\ref{eq12}) to the case where the condition of stability (\ref{eq3}) is not fulfilled, i.e., $\Lambda\neq 0 $. Let us show this.

Obvious analysis of Eqs.  (\ref{eq3}) and (\ref{eq10})-(\ref{eq13}) shows that with allowance for dimensions the expression:

\begin{equation}
\label{eq14}
\varepsilon=\frac{1}{2k}\Lambda=\frac{1}{2km}\partial_{i}\partial^{i}S,
\end{equation}

\noindent is, in general case, the variations of particle kinetic energy predetermined accordingly by variations of its momentum.

Now, if to add and simultaneously to subtract the complex expression ($i\varepsilon$) and to substitute also Eq. (\ref{eq4}) in the left-hand side of the Hamilton-Jacobi equation (\ref{eq7}), we get the generalized equation corresponding to the extended Eq. (\ref{eq8}):

\begin{equation}
\label{eq15}
  \begin{split}
    \frac{1}{2mk^2\psi }\partial_{i} \partial^{i}\psi - \frac{1}{2mk^2A}\partial_{i}\partial^{i}A - \frac{i}{2mk}\partial_{i}\partial^{i}S-\frac{1}{mk^2A}\partial_{i}A\left( \frac{1}{\psi }\partial^{i}\psi-\frac{1}{A}\partial^{i}A \right)-\\
    -\frac{1}{ikA\psi 
}\left[ {A\partial_{s}\psi-\psi \partial_{s}A} \right]-U-Q=0.
  \end{split}
\end{equation}

Repeating the ideology of derivation of Eq. (\ref{eq12}) it is easy from Eq .(\ref{eq15}) to obtain an equation of the "Stueckelberg" type, which is formally identical to Eq. (\ref{eq12}), but already under a more general condition imposed on the disturbing energy and the wave function:

\begin{equation}
\label{eq16}
Q = -\frac{1}{2mk^{2}A}\partial_{i}\partial^{i}A, 
\end{equation}

\begin{equation}
\label{eq17}
\partial_{s}A=-\frac{A}{2m}\partial_{i} \partial^{i}S - \partial_{i}A\partial^{i}S .
\end{equation}

Physical sense of (\ref{eq17}) for the wave function amplitude consists in the fact that with regard for the formulas for a classical velocity $v_{i} = \partial_{i} S / m $ and the probability density $P(q,t) = [A(q,t)]^2 $\linebreak
(the substantiation of this formula will be given below) in the configuration space it can be easily transformed into the equation of continuity, which represents the invariability of the total number of "particles" (phase points) or, in other words, the probability conservation law.

We now return to our problem of quantization on the basis of the simplest example. Let us consider a material point with mass $m$ in the field of conservative forces with the force function $U$, which depends, in the general case, on the evolution parameter. The problem on the stability of motions of such a point will be posed in the Cartesian coordinate system $q_{0}$, $q_{1}$, $q_{2}$, $q_{3}$. Denoting the momenta along the axes by $p_{0}$, $p_{1}$, $p_{2}$, $p_{3}$, respectively, we obtain known expression for the Hamiltonian::

\begin{equation}
\label{eq18}
K=\frac{1}{2m}p_{i}p^{i}.
\end{equation}

In this case, conditions (\ref{eq16}) and (\ref{eq17}) for the structure of disturbing and compensating forces admit the relations:
 
\begin{equation}
\label{eq19}
Q=-\frac{\hbar ^2}{2m}\frac{\partial_{i}\partial^{i}A}{A}{\begin{array}{*{20}c}
  , \hfill & {k=1/\hbar} \hfill \\
\end{array} },
\end{equation}

\begin{equation}
\label{eq20}
\partial_{s}A=-\frac{A}{2m}\sum {\frac{\partial^{2}S }{\partial q_i^{2} }}-\sum {\frac{\partial A}{\partial q_i }\frac{p^i }{m}},
\end{equation}

\noindent and the differential equation (\ref{eq12}) defining stable motions takes the form:

\begin{equation}
\label{eq21}
i\hbar \frac{\partial \psi }{\partial s}=-\frac{\hbar ^2}{2m}\partial_{i}\psi\partial^{i}\psi  
+U\psi .
\end{equation}

which coincides with the Stueckelberg equation. In our case, this equation restricts the choice of the integration constants in the full Hamilton-Jacobi integral. In what follows, we call Eq.~(\ref{eq21}) the Stueckelberg-Chetaev equation, by emphasizing the specific feature of its origin.

It is of interest to consider the case related to the inverse substitution of the wave function (\ref{eq4}) in the Schr\"{o}dinger equation (\ref{eq21}), that generates an equivalent system of equations known as the Bohm-Madelung system of equations \cite{bib-07,bib-08,bib-09}:

\begin{equation}
\label{eq22}
\partial_{s}A=-\frac{1}{2m}\left[A\partial_{i}\partial^{i}S+2\partial_{i}A\partial^{i}S \right],
\end{equation}

\begin{equation}
\label{eq23}
\partial_{s}S=-\left[ {\frac{1}{2m}\partial_{i}S\partial^{i}S+U-\frac{\hbar 
^2}{2m}\frac{\partial_{i}\partial^{i} A}{A}} \right].
\end{equation}

It is very important that the last term in Eq.~(\ref{eq23}), which is the "quantum" potential of the so-called Bohm $ \psi $-field \cite{bib-07,bib-08,bib-09,bib-23} in the interpretation of Ref. \cite{bib-07}, coincides exactly with the dissipation energy $Q$ in (\ref{eq19}). At the same time, Eq.~(\ref{eq22}) is identical to the condition for $\partial A$/$\partial s $ in (\ref{eq17}) and (\ref{eq20}).

If we make substitution of the type:

\begin{equation}
\label{eq24}
P(q,s)= \psi \psi^{*}=[A(q,s)]^{2},   
\end{equation}

\noindent then Eqs.~(\ref{eq22}) and (\ref{eq23}) can be rewritten in the form:

\begin{equation}
\label{eq25}
\frac{\partial P}{\partial s}=-\frac{1}{m}\partial_{i}(P\cdot \partial^{i} S),
\end{equation}

\begin{equation}
\label{eq26}
\partial_{s}S+\frac{\partial_{i}S\partial^{i}S}{2m}+U-\frac{\hbar 
^2}{4m}\left[ {\frac{\partial_{i}\partial^{i} P}{P}-\frac{1}{2}\frac{\partial_{i}P\partial^{i}P}{P^2}} 
\right]=0.
\end{equation}

Here, Eq.~(\ref{eq25}) has a clear physical sense: $P(q, s)$ is the probability density to find a particle in a certain place of the 4-space, and $ \partial_{i} S / m $ is, according to (\ref{eq13}), the  4-velocity of this particle. In other words, Eq.~(\ref{eq25}) is nothing but the equation of continuity which indicates how the probability density $P(q, s)$ moves according to the laws of classical mechanics with a classical velocity at every point.

On the other hand, we can show that $P(q,s)$ is also the probability density function for the number of particle trajectories, that is substantiated in the following way. We assume that the influence of disturbing forces generated by the potential $Q$ on a wave packet at an arbitrary point of the configuration space is proportional to the density of particle trajectories ($\psi \psi^{*} = A^{2}$) at this point. This implies that the disturbing forces do not practically perturb the packet, if the relation

\begin{equation}
\label{eq27}
\int {Q\psi \psi ^\ast dV\Rightarrow \min}, ~~~\int {\psi \psi ^\ast dV=1} ,
\end{equation}

\noindent is satisfied, where $dV$ stands for an element of the configuration space volume. This means, in its turn, that the disturbing forces admit the absolute stability on the whole set of motions in the configuration space only if condition (\ref{eq27}) is satisfied or, in other words, if the following obvious condition of the equivalent variational problem (for $Q$) fulfills:

\begin{equation}
\label{eq28}
\delta\int {Q\psi \psi ^\ast dV} =\delta Q =0.
\end{equation}

The variational principle (\ref{eq28}) is, in essence, the principle of least action of a perturbation.Below, we call it the principle of least action of perturbations by Chetaev \cite{bib-01}. Using the previous notations, we can write the following equality for $Q$:

\begin{equation}
\label{eq29}
Q=-\partial_{s}S-U-T=-\frac{\partial S}{\partial 
t}-U-\frac{1}{2}\partial_{i}S\partial^{i}S .
\end{equation}

On the other hand, if (\ref{eq4}) holds true, it is easy to show that:

\begin{equation}
\label{eq30}
\frac{1}{2m}\partial_{i}S\partial^{i}S=-\frac{1}{2mk^2\psi ^2}\partial_{i}\psi\partial^{i}\psi+\frac{1}{2mk^2A^2}\partial_{i}A\partial^{i}A + 
ik\frac{1}{2mk^2A^2}\partial_{i}A\partial^{i}S .
\end{equation}

Then it is necessary to perform the following subsequent substitutions. First, we substitute relation (\ref{eq30}) in (\ref{eq29}) and then introduce the result in the equation corresponding to the variational principle (\ref{eq28}).

As a result of the indicated procedure of substitutions, we obtain the relation which is exactly equal
to Eq. (\ref{eq15}), i.e., to the extended equation (with regard for (\ref{eq3}), (\ref{eq14}), (\ref{eq15})] of type (\ref{eq8}). Hence, the structural
expressions and the necessary condition of stability which follow from it coincide with (\ref{eq16}), (\ref{eq17}),
and (\ref{eq12}), respectively. This means that, on the basis of the Chetaev variational principle (\ref{eq28}), we get
the independent confirmation of the fact that the physical nature of P(q, t) reflects really not only the
traditional notion of the probability density for a particle to be at a certain place of the space according
to the Bohm-Madelung equation of continuity (\ref{eq25}) but plays also the role adequate to that of the
probability density of the number of particle trajectories.

\section{Neutrino oscillation measurements as verifiability criterion
of different concepts of temporal evolution}

Here the natural question raises: "What of the two equations realizing ideology of standard or parametrized RQM is substantiated and valid within the framework of modern experimental relativistic dynamics or, in other words, what of the two concepts of temporal evolution reflects nature of physical reality of our world more sufficiently?" It is interesting, that one of variants of the theoretical formulation of such a problem is stated for the first time by Fanchi \cite{bib-17,bib-22}. Because we will use the results of this work below, let us briefly consider the formulation of this problem.

Following \cite{bib-22}, we consider the temporal model of vacuum-flavor mixing neutrino within the framework of the standard and pRQM models. At first, this is caused by the fact that, according to \cite{bib-22}, differences between the standard and pRQM models are highlighted by developing a general formalism for vacuum-flavor mixing and then applying the formalism within the context of each theory. Secondly, the resulting two expressions for the probability of transition from one neutrino flavor to another are experimentally testable that is the one of purposes of our research, which is given in next sections.

Considering the temporal model of vacuum-flavor mixing, we imply the case of two-state mixing, i.e., we confine ourself to consideration of the two mass states  $\left\lbrace \mid \nu_{j} \right\rangle \}$ and two neutrino flavor states  $\left\lbrace \mid \nu_{\alpha} \right\rangle \}$, which may be written as the two-component column vectors:

\begin{equation}
\label{eq31} 
\left\{{\left| \nu _{j}  \right\rangle} \right\}=\left[\begin{array}{c} {{\left| \nu _{1}  \right\rangle} } \\ {{\left| \nu _{2}  \right\rangle} } \end{array}\right]\begin{array}{cc} {,} & {\left\{{\left| \nu _{\alpha }  \right\rangle} \right\}} \end{array}=\left[\begin{array}{c} {{\left| \nu _{e}  \right\rangle} } \\ {{\left| \nu _{\mu }  \right\rangle} } \end{array}\right].                                                
\end{equation} 
\textbf{}

The temporal evolution equation for state may be written in terms of the temporal evolution operator \textbf{T} as:

\begin{equation}
\label{eq32}    
\textbf{T${\left| \nu _{j}  \right\rangle} =i\hbar \frac{\partial }{\partial \tau } {\left| \nu _{j}  \right\rangle} =$T${}_{j}$ ${}_{{\left| \nu _{j}  \right\rangle} }$} 
\end{equation}

\noindent with the formal solution

\begin{equation}
\label{eq33}
\left[ \begin{array}{c} {{\left| \nu _{1} (\tau ) \right\rangle} } \\ 
                        {{\left| \nu _{2} (\tau ) \right\rangle} } 
       \end{array} \right] = 
\left[ \begin{array}{cc} \exp (-i\frac{T_{1} \tau }{\hbar } ) & 0 \\
                         0 & \exp (-i\frac{T_{2} \tau }{\hbar } )
       \end{array} \right]
\left[ \begin{array}{c} \left| \nu _{1} (0) \right\rangle \\ 
                        \left| \nu _{2} (0) \right\rangle \end{array} \right].                                         
\end{equation} 

It is obvious \cite{bib-22}, that the eigenvalue \textbf{T${}_{j}$} of the temporal evolution operator \textbf{T} depends on the context of theory. In the standard theory, \textbf{T${}_{j}$}  is energy ${E}_{j}$ of state $j$, while in pRQM it is the eigenvalue ${\textbf{K}}_{j}$ of the mass operator for mass state $j$. Explicit free-particle expressions for the eigenvalues are:

\begin{equation} 
\label{eq34}
(\textbf{T}_{j})_{Standard} =E_{j} =\left[p^{2} c^{2} +m_{j}^{2} c^{4} \right]^{{1 \mathord{\left/{\vphantom{1 2}}\right.\kern-\nulldelimiterspace} 2} }, \\  \tau _{Standard} =t,                                              
\end{equation} 

\noindent and

\begin{equation}
  \begin{split}
    (\textbf{T}_{j})_{pPQM} &= K_{j} = \frac{\hbar ^{2} }{2m_{j}} k_{j}^{\mu} k_{j\mu} =\\
    & = \frac{\hbar ^{2} }{2m_{j} } \left[(\omega _{j} / c^{2})-\vec{k}_{j} \cdot \vec{k}_{j} \right], ~~~(\tau )_{pRQM} = s.
    \end{split}
  \label{eq35} 
\end{equation} 

Here  $\left\{k_{j}^{\mu } \right\}=\left\{\omega _{j} ,\vec{k}_{j} \right\}$ denotes the energy-momentum four-vector in laboratory frame and $m_{j}$ is the mass of state for a metric with signature $\left\lbrace +,-,-,-\right\rbrace $. According \cite{bib-22}, in pRQM, all four component of the energy-momentum four-vector are observable and $m_{j}$  is function of statistical values of the energy-momentum four-vector. In this case the energy $\omega_{j}$ of state $j$ is treated in pRQM as an independent variable and does not satisfy an equation like (\ref{eq34}), except in a statistical sense. By contrast $m_{j}$ is an input parameter in the standard model. In this case the standard picture of neutrino oscillation by flavor mixing in vacuum assumes that the three-momentum is constant for all states and has magnitude $p^{2} = h^{2} \vec{k}\cdot \vec{k}$.

It is known that a neutrino oscillation process is descried by hypothesizing that the mass basis $\mid \nu_{j} \rangle$ is related to the flavor basis $\mid \nu_{\alpha} \rangle$ by a unitary transformation $U$ such that

\begin{equation} 
  \label{eq36}
\left\{{\left| \nu _{\alpha }  \right\rangle} \right\}=U\left\{{\left| \nu _{j}  \right\rangle} \right\}\equiv \left[\begin{array}{cc} {\begin{array}{c} {\cos \theta } \\ {-\sin \theta } \end{array}} & {\begin{array}{c} {\sin \theta } \\ {\cos \theta } \end{array}} \end{array}\right]\left\{{\left| \nu _{j}  \right\rangle} \right\} , 
\end{equation}
 
\noindent and $\theta$ is the mixing angle of mass states in vacuum. For two flavors of neutrinos and the unitary transformation $U$ we have the $\tau$-dependent state:

\begin{equation} 
  \label{eq37} 
{\left| \nu _{e} (\tau ) \right\rangle} =e^{{-iT_{1} \tau  \mathord{\left/{\vphantom{-iT_{1} \tau  \hbar }}\right.\kern-\nulldelimiterspace} \hbar } } \cos \theta {\left| \nu _{1}  \right\rangle} +e^{{-iT_{2} \tau  \mathord{\left/{\vphantom{-iT_{2} \tau  \hbar }}\right.\kern-\nulldelimiterspace} \hbar } } \sin \theta {\left| \nu _{2}  \right\rangle} .                                     
\end{equation} 

The probability of transition from $\nu_{e} (\tau)$ to $\nu_{\mu}$ is the square of the magnitude of the transition matrix element

\begin{equation} 
  \label{eq38} 
P\left(\nu _{e} \to \nu _{\mu } \right)=\left|\left\langle \nu _{\mu } {\left| \nu _{e} (\tau ) \right\rangle} \right. \right|^{2} =\sin ^{2} 2\theta \sin ^{2} \frac{(T_{2} -T_{1} )\tau }{2\hbar }.                              
\end{equation}

\noindent Thus, the survival probability $P (\nu_{e} \rightarrow \nu_{e})$ has the form

\begin{equation} 
\label{eq39} 
P\left(\nu _{e} \to \nu _{e} \right)=1-P(\nu _{e} \to \nu _{\mu } )=1-\sin ^{2} 2\theta \sin ^{2} \frac{(T_{2} -T_{1} )\tau }{2\hbar }.                         
\end{equation} 

It is obvious that, that standard model and pRQM model according to (\ref{eq34}-\ref{eq35}) differ in the explicit forms of $T_{j}$ and $\tau$. Therefore, the expression for the survival probability $P (\nu_{e} \rightarrow \nu_{e})$, obtained in the framework of standard model and pRQM model, will have the next form:

\begin{equation}
  \begin{split}
    P_{S\tan dard} (\nu _{e} \to \nu _{e} ) &= 1-\sin ^{2} 2\theta \sin ^{2} \left[ \frac{(m_{2}^{2} -m_{1}^{2} )^{2} c^{4} }{4\hbar E_{v} } \frac{L}{c} \right] =\\
    & =1-\sin ^{2} 2\theta \sin ^{2} \alpha _{S\tan dard},
  \end{split}
   \label{eq40} 
\end{equation}

\noindent where the coordinate time $t$ is replaced in standard model by the distance $L$ from the neutrino source to the detector, so that $t = L / c$. Note that for ultra-relativistic neutrino it is assumed that  that $p_{1} \approx p_{2} = p$, and that $p / c \gg m_{2}, m_{1}$, and $p \approx E_{\nu} / c$, where $E_{\nu} = m_{1} c^{2} / (1 - \beta^{2})^{1/2}$ is the energy of the incident neutrino ($\beta = \upsilon / c$):

\begin{equation}
  \begin{split}
    P_{pRQM} (\nu _{e} \to \nu _{e} ) & = 1-\sin ^{2} 2\theta \sin ^{2} \left[\frac{(m_{2} -m_{1} )c^{2} }{4\hbar } \frac{L}{c} \frac{(1-\beta ^{2} )^{{1 \mathord{\left/{\vphantom{1 2}}\right.\kern-\nulldelimiterspace} 2} } }{\beta } \right] = \\
    &= 1-\sin ^{2} 2\theta \sin ^{2} \left[\frac{(m_{2}^{2} -m_{1}^{2} )c^{4} }{4\hbar E_{\nu } } \frac{L}{c} \frac{m_{1} }{m_{1} +m_{2} } \frac{1}{\beta } \right]=1-\sin ^{2} 2\theta \sin ^{2} \alpha _{pRQM}.
  \end{split}
\label{eq41} 
\end{equation} 

Based on well-known experiments on solar and reactor neutrino detection \cite{bib-24} one can to conclude that existence of neutrino oscillations is indisputably proved and to assume without loss of generality that the mass difference between neutrino flavor states is very small, i.e., $m_{1} \approx m_{2}$. Thus, it is easy to see that in the case of ultra-relativistic neutrino (i.e., $\beta \rightarrow 1$) the dynamical factors $\alpha_{Standard}$ and $\alpha_{pRQM}$ are connected by the following simple expression: 

\begin{equation} 
  \label{eq42} 
\alpha _{pRQM} \cong 2\alpha _{Standard}.                                                        
\end{equation}

\noindent or

\begin{equation} 
  \label{eq43}
\left(\Delta m_{21}^{2} \right)_{pRQM} \cong 2\left(\Delta m_{21}^{2} \right)_{Standard},                                            
\end{equation} 

\noindent where $\Delta m_{21}^{2} = m_{2}^{2} - m_{1}^{2}$.

In other words, the ratio of dynamical factors does not equal to 1 because of the difference in operators $K_{j}$ between standard paradigm and pRQM. It was pointed out in Fanchi \cite{bib-22} that the pRQM mass operator is the direct result of the probabilistic basis of the theory. Consequently  according to \cite{bib-22}, the search for massive neutrinos is a test not only of the flavor-mixing hypothesis, but also the validity of standard model and pRQM.

\section{Neutrino mass matrix via see-saw mechanism}

It is obvious, that to analyse a general formalism for vacuum-flavor mixing of neutrino within the context of the standard and pRQM models we need the estimations of absolute value for neutrino mass based on modern experimental data for solar and atmospheric neutrinos. With that end in view we consider below the problem of determining the neutrino masses by use of special method for construction of neutrino mass matrix, which was described for the first time in \cite{bib-25}, and advantages of the mass-squared differences and mixing parameters obtained from neutrino oscillation experiments.

It is known that one of the interesting mechanisms which can generate a small neutrino mass is the see-saw mechanism, in which the right-handed neutrino $\nu_{R}$ has a large Majorana mass $M_{N}$ and the left-handed neutrino $\nu_{L}$ obtain a mass through leakage of the order of ($m / M_{N}$) with $m$ is the Dirac mass. According to see-saw mechanism \cite{bib-26}, the neutrino mass matrix $M_{\nu}$ is given by:

\[M_{\nu } \approx -M_{D} M_{N}^{-1} M_{D}^{T} ,\]

\noindent where $M_{D}$ and $M_{N}$ are the Dirac and Majorana mass matrix respectively.

In order to obtain neutrino mass matrix $M_{\nu}$ that can give correct predictions on mass-squared differences and mixing parameters, Damanik \cite{bib-25} modified the the neutrino mass matrix $M_{\nu}$ by introducing one parameter $\delta$ to perturb the diagonal elements of $M_{\nu}$. In the way that the perturbed mass matrix satisfies the requirement $Tr \left(M_{\nu}\right) \equiv 3 P$. At the same time, the neutrino mass matrix $M_{\nu}$ can be written in the following form \cite{bib-25}:

\begin{equation} 
  \label{eq44} 
M_{\nu } = \left( \begin{array}{ccc} P+2 \delta & Q & Q \\
           Q & P-\delta & Q \\ 
           Q & Q & P-\delta \end{array} \right).                                                    
\end{equation} 

The eigenvalues of the neutrino mass matrix $M_{\nu}$ ${}_{ }$ in (\ref{eq44}) are:

\begin{equation} 
\label{eq45} 
\beta _{1,2} =P+\frac{Q}{2} +\frac{\delta }{2} \mp \frac{\sqrt{9\delta ^{2} -6Q\delta +9Q^{2} } }{2},                                       
\end{equation} 

\begin{equation}
\label{eq46} 
  \beta_{3} =P-Q-\delta .
\end{equation} 

If the neutrino mass matrices $M_{\nu}$ in (\ref{eq44}) is diagonalized by mixing matrix $V$ in the equation

\begin{equation} 
\label{eq47}
 \left( \begin{array}{c} \nu_{e} \\ \nu_{\mu} \\ \nu_{\tau} \end{array} \right) = V \left( \begin{array}{c} \nu_{1} \\ \nu_{2} \\ \nu_{3} \end{array} \right),                                                            
\end{equation}
 
\noindent where

\begin{equation} 
\label{eq48} 
V = \left( \begin{array}{ccc} \cos \theta & -\sin \theta & 0\\ 
    \sin \theta / \sqrt{2} & \cos \theta / \sqrt{2} & -1 / \sqrt{2} \\ 
    \sin \theta / \sqrt{2} & \cos \theta / \sqrt{2} & 1 / \sqrt{2}
\end{array} \right),                                         
\end{equation}
 
\noindent then we obtain

\begin{equation} 
\label{eq49} 
\tan ^{2} (2\theta )=\frac{8Q^{2} }{(Q-3\delta )^{2} },                                                     
\end{equation}
 
\noindent and neutrino mass as follow:

\begin{equation} 
\label{eq50} 
m_{1} = P + \frac{Q}{2} +\frac{\delta }{2} -\frac{\sqrt{9\delta ^{2} -6Q\delta +9Q^{2} } }{2},                                        
\end{equation} 

\begin{equation} 
\label{eq51} 
m_{2} = P+\frac{Q}{2} +\frac{\delta }{2} +\frac{\sqrt{9\delta ^{2} -6Q\delta +9Q^{2} } }{2},                                        
\end{equation} 

\begin{equation} 
\label{eq52} 
m_{3} =P-Q-\delta .                                                           
\end{equation} 

One can see that the obtained neutrino masses in Damanik's scenario \cite{bib-25} is an inverted hierarchy with masses: $\mid m_{3} \mid < \mid m_{1} \mid < \mid m_{2} \mid$.

To solve the system of the equations ~(\ref{eq49})-(\ref{eq52}) with respect to the values of neutrino masses it is necessary to supplement this system with the known values of mass-squared differences and mixing parameters, which obtained from neutrino oscillation experiments. From the recent global analysis of the neutrino oscillation data we refer to the following best-fit values of solar neutrino mass-squared differences \cite{bib-27}:
 
\begin{equation} 
\label{eq53} 
\Delta m_{21}^{2} =(7.5_{-0.20}^{+0.19} )\times 10^{-5} ~~eV^{2}, 
\end{equation}

\noindent with

\begin{equation} 
\label{eq54} 
\tan ^{2} \theta _{12} =0.452_{-0.033}^{+0.035},                                                      
\end{equation} 

\noindent and for the atmospheric neutrino mass-squared differences \cite{bib-28}:

\begin{equation} 
\label{eq55} 
\Delta m_{32}^{2} =(2.32_{-0.08}^{+0.12} )\times 10^{-3} ~~eV^{2}.
\end{equation} 

If $\theta$ is the $\theta_{12}$ in (\ref{eq54}), we have  from (\ref{eq49}) we have $\delta = -0.050909 ~Q$. If we insert this $\delta$ value into (\ref{eq50}-\ref{eq52}) then we obtain the neutrino mass as follow:

\begin{equation} 
\label{eq56}
m_{1} = P - 1.052610 ~Q,
\end{equation} 

\begin{equation} 
\label{eq57}
m_{2} = P + 2.001700 ~Q,
\end{equation} 

\begin{equation} 
\label{eq58}
m_{3} = P - 0.949910 ~Q.                                                      
\end{equation} 

Using the advantages of experimental data of neutrino oscillations in Eqs. (\ref{eq53}) and (\ref{eq55}) from Eqs.(\ref{eq56},\ref{eq57}), we have the neutrino masses Table~\ref{tab-1}) obtained within the framework of standard model and pRQM model. 

\begin{table}[h]
\caption{\label{tab-1}Neutrino mass $\left|m_{i} \right|$ as function of dynamical factors $\alpha_{Standard}$ and $\alpha_{pRQM}$.} 

\begin{center}
\lineup
\begin{tabular}{*{3}{l}}
\br                              
& $\alpha_{Standard}$ & $\alpha_{pRQM}$\cr 
\mr
$\Delta m_{21}^{2}\times 10^{5} ~~eV^{2}$ & 7.50 & 15.00 \cr
$\Delta m_{32}^{2}\times 10^{3} ~~eV^{2}$ & 2.32 & 4.64 \cr
\mr 
$tg^{2} \theta$ & 0.452 & 0.452 \cr 
\mr
$\left|m_{1} \right|$ & 0.130955 & 0.185056 \cr 
$\left|m_{2} \right|$ & 0.131141 & 0.185461 \cr 
$\left|m_{3} \right|$ & 0.121975 & 0.172499 \cr 
\br
\end{tabular}
\end{center}
\end{table}

Thus, we estimated the neutrino masses within the framework of both models. So long as these estimations satisfy the known restriction on the total neutrino mass:

\begin{equation} 
\label{eq59}
\sum m_{a}\leq 0.6 eV, ~~~ \nu_{a}=\nu_{e},\nu_{\mu},\nu_{\tau},
\end{equation}

\noindent it is impossible to answer  question, what of the two representations realizing the ideology of standard or parametrized RQM reflects the nature of physical reality of our world more sufficiently. A valid answer to this question can be given only when the experimental values of the neutrino masses will be obtained. 

Below we will attempt to solve this problem even if at qualitative level using modern cosmological observations.

\section{The dimension of the neutrino cloud and large-scale structure of Universe}

In this section we compare the theoretical value of void structure of Universe obtained by use of foregoing alternative values of neutrino masses with the experimental value of average size of large-scale structure of Universe obtained on the basis of analysis of so-called the matter power spectrum of density fluctuation in space-time. The main purpose of such a comparison is the reveal of dominant role one of the two physical concepts of temporal evolution, which are considered above. 

So, according to Muraki \cite{bib-29,bib-30}, the possible role of massive neutrinos in our Universe is as follows. Massive neutrinos are attracted by the gravitational force. However, because of the degeneracy force, they will not collapse to form a supper black hall. The Jeans mass of neutrinos grows and fluctuation of the baryonic matter will be erased. The fluctuation of the baryonic matter can grow only at the border between neutrino bubbles. Consequently, it is possible to assume that the observed void structure of the Universe was formed by the neutrino degenerate force and grows of the density fluctuation at the border between voids.

Using the analogy with neutron star, whose radius \footnote{The canonical radius of a neutron star \cite{bib-31} can be obtained by the balance between the degenerate force of neutrons and the gravitational force \cite{bib-29,bib-30}: $r_{n}=1.2(\hbar^{2}/G)(1/M_{n}) 1/3 (1/m_{n})^{8/3} \simeq 10$ km, where $M_{n}$ corresponds to a 1.4 solar mass.} can be obtained by the balance between the degenerate force of neutrons and the gravitational force, Muraki \cite{bib-29,bib-30} in bounds of his script(scenario) got the next expression for the radius of a neutrino cloud:
 
\begin{equation} 
\label{eq60} 
r_{\nu } = 1.2 \left( \frac{\hbar ^{2} }{G} \right) \left( \frac{1}{M_{\nu}} \right) ^{1/3} \left( \frac{1}{m_{\nu}} \right) ^{8/3},
\end{equation}

\noindent where the neutrino cloud mass is

\begin{equation} 
\label{eq61} 
M_{\nu } =\left({4 / 3} \right) \pi r_{\nu }^{3}\times 110\times 10^{6} \left[m^{3} \right]\times m_{\nu } \left[eV\right]\times 1.7\times 10^{-36} \left[kg\right].                           
\end{equation} 

However, in this scenario it is necessary to take into account also the effect of baryonic matter and dark matter \cite{bib-29,bib-30}. 
According to the last analysis of the WMAP7 data, total matter fraction  $\Omega_{M}$ is $\Omega_{M} = 26.6 \pm 2.9 \%$ \cite{bib-32}, while the energy density of all types of neutrinos is  $\Omega_{\nu} \leq 0.014$ \cite{bib-33} \footnote{In terms of neutrino mass this restriction ($\Omega_{\nu}\leq0.014$) means that $\sum m_{\nu}\leq0.6 ~~eV$ \cite{bib-24,bib-33}, which is satisfied by the values of triplet neutrino masses listed in Table~\ref{tab-1}.}. That means that the mass of (dark matter + hydrogen gas) is 19 times heavier than the mass of neutrinos. Therefore if we replace $110 m_{\nu}$ by $2090 m_{\nu}$ and at the same time take into account  the contribution to the gravitational mass from other neutrinos, we will obtain modern expression of Muraki relation for the neutrino cloud diameter:

\begin{equation} 
\label{eq62} 
d_{\nu } = 7.1 \left[Mpc\right]\times m_{\nu }^{-3/2} \left[eV\right].                                                   
\end{equation} 

Using this expression it is easy to estimate the average diameter of neutrino cloud for different neutrino masses predetermined by the corresponding dynamical factors  $\alpha_{Standard}$ and $\alpha_{pRQM}$ (see Table~\ref{tab-1}):

\begin{equation} 
\label{eq63} 
m_{\nu }^{pRQM} =0.185461eV \to d_{\nu }^{pRQM} \cong 90 Mpc,                              
\end{equation} 

\begin{equation} 
\label{eq64} 
m_{\nu }^{Standard} = 0.131141 ~eV \to d_{\nu }^{Standard}\cong 150 Mpc.                           
\end{equation} 

\begin{figure}
\begin{center}
\includegraphics*[width=3.7in]{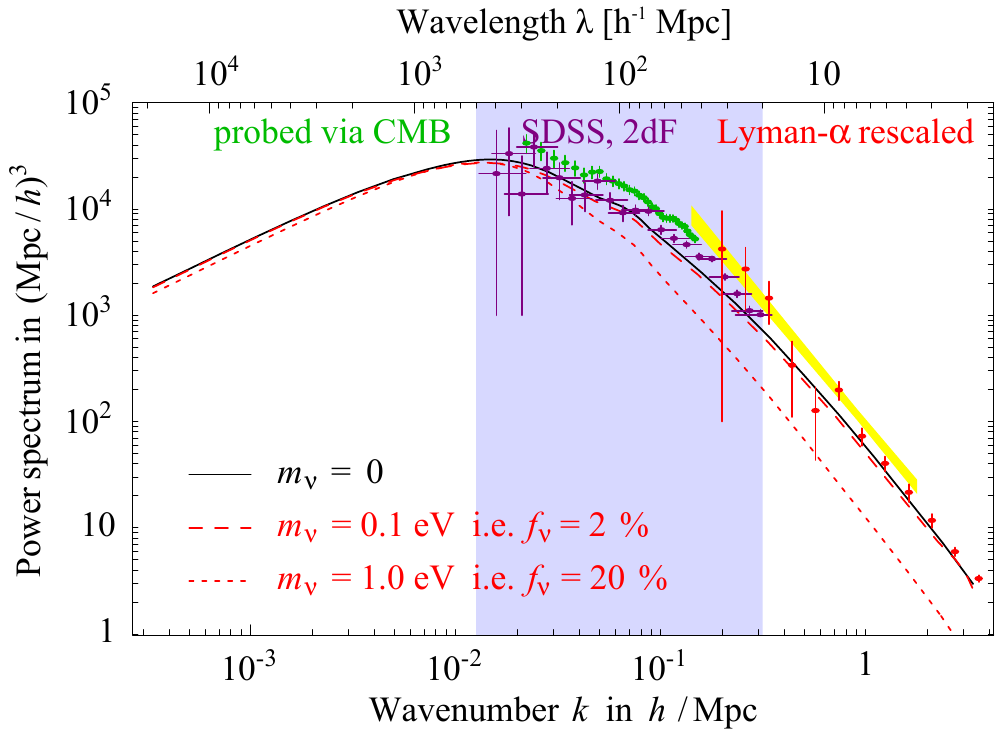}
\end{center}
\caption{\label{fig-1} The matter power spectrum $P(k)$ predicted by the best-fit $\Lambda$CDM cosmological model (continuous curve) and how neutrino masses affect it (dashed curves). Measurements at different scales have been performed with different techniques, that slightly overlap. The data points do not show the overall uncertainty that plagues galaxy surveys (SDSS, 2dF) at intermediate scales and especially Lyman-$\alpha$ data at smaller scales i.e. at larger $k$. Here $f_{\nu} \equiv \rho_{\nu} / \rho_{DM}$, where $\rho_{\nu}$ -- neutrino density, $\rho_{DM}$ -- density of dark matter. The wide vertical stripe identifies an interval maximum best-fit $\Lambda$CDM cosmological model (continuous curve), corresponding to the range of wavelength values $\lambda = 20 \div 500 ~[{h}/ Mpc]$. Adopted from \cite{bib-24}.}
\end{figure}

On the other hand, it is known, that the average size of typical observed void structure (dark matter + hydrogen gas), which plays the role of characteristic dimension of large-scale structure of the Universe, are predetermined by so-called of the matter power spectrum of density fluctuation in space-time (Fig.~\ref{fig-1}).

It is easy to show (see Figure~\ref{fig-1} \cite{bib-24}), that the range of large-scale structure (LSS) set by the range of the wavelength $\lambda = 20 \div 500  ~[{h}^{-1}Mpc]$ ($h = 0.704$ \cite{bib-32}) is described by the average characteristic dimension of LSS of the Universe:

\begin{equation} 
\label{eq65}
  \left\langle \lambda _{LSS} \right\rangle \sim 90 ~~Mpc.
\end{equation}
 
\noindent This is confirmed also by the known results of Tegmark \textit{et al.} \cite{bib-35}.

Comparative analysis of expressions (\ref{eq65}) and (\ref{eq63},\ref{eq64}) shows that the characteristic dimensions of LSS (\ref{eq65}) and neutrino cloud (\ref{eq63}) practically coincide:

\begin{equation} 
\label{eq66}
  \left\langle \lambda _{LSS} \right\rangle \sim d_{\nu }^{pRQM}.                                                           
\end{equation} 

This result shows that the physical concept of temporal evolution described by Stueckelberg equation is more adequate. 

\section{Conclusion}

Certainly, it is impossible to consider obtained result as a proof, but it is a strong argument in favour of the Stueckelberg concept of temporal evolution, which reflects the nature of physical reality of our world more sufficiently than the standard (Dirac) concept. It is clear that direct proof in favour of one or the other physical concept of temporal evolution will be possible only when the experimental value of neutrino mass will be determined. 

In this sense, it is appropriate to note here that "... time and again a breakthrough in the field of new physics was related to neutrino. The first theory of $\beta$-decay proposed by Fermi was based on the Pauli hypothesis of existence of neutrino. Phenomenological (V-A) theory of weak interactions has arisen from the two component neutrino theory considered by Landau, Lee, Yang and Salam. Evidence in favour of the Standard Model of weak interaction of Glashow, Weinberg and Salam had been received on a neutrino beam (discovery of neutral currents)" \cite{bib-36}. At the present time, we observe a new phenomenon - neutrino oscillations, which result from temporal evolution of massive neutrinos and indicate new physics beyond the Standard Model. Undoubtedly, that discovery of neutrino mass in the near future (for example, by research of high-energy edge of the tritium $\beta$-spectrum) can become the basis for reconsideration not only of the key issues of quantum theory, but a holistic understanding of the nature of physical reality.

\smallskip

\end{document}